\documentclass[lettersize,journal]{IEEEtran}
\usepackage{amsmath,amsfonts}
\usepackage{algorithmic}
\usepackage{algorithm}
\usepackage{array}
\usepackage{textcomp}
\usepackage{stfloats}
\usepackage{url}
\usepackage{verbatim}
\usepackage{graphicx}
\usepackage{booktabs}
\usepackage{subcaption}
\usepackage{cite}
\begin{document}

\title{EDA-Q: Electronic Design Automation for Superconducting  Quantum Chip}

\author{Bo Zhao$^1$$^{\dagger}$, Zhihang Li$^1$$^{\dagger}$, Xiaohan Yu$^1$$^\dagger$, Benzheng Yuan$^1$, Chaojie Zhang$^1$, Yimin Gao$^1$, Weilong Wang$^1$, Qing Mu$^1$, Shuya Wang$^1$, Huihui Sun$^1$, Tian Yang$^1$, Mengfan Zhang$^1$, Chuanbing Han$^1$, Peng xu$^1$, Wenqing Wang$^1$ and Zheng Shan$^1$$^\star$\\
\textit{\small{$^\star$ Corresponding author: shanzhengzz@163.com}}; \\
\textit{\small{$^\dagger$ These authors contributed equally to this work}};\\
\textit{\small{$^1$ Laboratory for Advanced Computing and Intelligence Engineering}}
}



\maketitle

\begin{abstract}
Electronic Design Automation (EDA) plays a crucial role in classical chip design and significantly influences the development of quantum chip design. However, traditional EDA tools cannot be directly applied to quantum chip design due to vast differences compared to the classical realm. Several EDA products tailored for quantum chip design currently exist, yet they only cover partial stages of the quantum chip design process instead of offering a fully comprehensive solution. Additionally, they often encounter issues such as limited automation, steep learning curves, challenges in integrating with actual fabrication processes, and difficulties in expanding functionality. To address these issues, we developed a full-stack EDA tool specifically for quantum chip design, called EDA-Q. The design workflow incorporates functionalities present in existing quantum EDA tools while supplementing critical design stages such as device mapping and fabrication process mapping, which users expect. EDA-Q utilizes a unique architecture to achieve exceptional scalability and flexibility. The integrated design mode guarantees algorithm compatibility with different chip components, while employing a specialized interactive processing mode to offer users a straightforward and adaptable command interface. Application examples demonstrate that EDA-Q significantly reduces chip design cycles, enhances automation levels, and decreases the time required for manual intervention. Multiple rounds of testing on the designed chip have validated the effectiveness of EDA-Q in practical applications.
\end{abstract}


\section{Introduction}
\begin{table*}[h]
	\caption{The Comparision of Different EDA Tools for Quantum Chip}
	\label{The Comparision of Different EDA Tools for Quantum Chip}
	\begin{tabular}{ccccccc}
		\toprule
		\ \ &EDA-Q&Qiskit-Metal&QuantumPro&KQCircuit&SpinQ&Origin Unit\\
		\midrule
		\ Topology Design                       &\checkmark &×&×&×&×&×\\
		\ Equivalent Circuit Design             &\checkmark &×&×&×&×&\checkmark \\
		\ GDS Layout Design                     &\checkmark &\checkmark &\checkmark &\checkmark &\checkmark &\checkmark \\
		\ Device Mapping                        &\checkmark &×&×&×&×&×\\
		\ Routing Design                        &\checkmark &\checkmark &\checkmark &\checkmark &\checkmark &\checkmark \\
		\ Fabrication Process Mapping           &\checkmark &×&×&×&×&\checkmark \\
		\ Simulation Verification\&Optimization &\checkmark &\checkmark &\checkmark &×&×&\checkmark \\
		\bottomrule
	\end{tabular}
	\centering
\end{table*}
Quantum computing has significant potential for efficiently solving specific problems in areas such as machine learning\cite{HOUSSEIN2022116512}\cite{Huang2020PowerOD}\cite{Caro2021GeneralizationIQ}, cryptography\cite{shor1999polynomial}\cite{Kumar2021StateoftheArtSO}, and chemistry\cite{Motta2021EmergingQC} due to its inherent quantum superposition and entanglement properties. There are different ways to physically construct quantum computers,  and one particular method called superconducting quantum computing is getting more and more attention because it has been consistently improving in terms of gate and measurement accuracy\cite{article}\cite{inproceedings}. In the era of noisy intermediate-scale quantum computing\cite{Preskill2018}, despite its demonstrated potential for exponentially accelerating the solution of certain problems compared to classical supercomputers, there remain substantial challenges arising from limited coherence times and diverse sources of noise. High-quality design and fabrication of quantum chips are crucial foundations of the quantum ecosystem. A comprehensive chip design process typically comprises several stages including design, fabrication, thorough testing, and continuous optimization, which are time-consuming and costly. Consequently, the need to reduce the design cycle and improve chip design quality has become a pressing matter that needs to be addressed.

EDA is crucial in classical chip design as it plays essential roles in high-level abstraction, automated design, process optimization, and simulation verification. Similarly, the creation of quantum chips involves a series of complex procedures. There are currently multiple products available for use in the field of Quantum Electronic Design Automation (QEDA). Qiskit Metal\cite{qiskit-metal} is an open-source software development kit (SDK) that specifically focuses on designing superconducting quantum circuits and computational devices. It is an essential part of the IBM Qiskit ecosystem. QuantumPro\cite{quantumpro}, represents a comprehensive and tailored process for creating superconducting quantum chips. Its PathWave platform integrates five fundamental features, including schematic design, layout creation, electromagnetic (EM) analysis, nonlinear circuit simulation, and quantum parameter extraction. KQCircuits\cite{KQCircuits}, created by IQM, employ a library of superconducting quantum circuit components that is built upon KLayout. This platform provides interfaces with simulation software, aiming to reduce the workload for designers during the quantum processor design process. SpinQ's Tianyi EDA tool\cite{SpinQ} expedites chip layout design by leveraging its component library and advanced automated routing algorithms. The Origin Unit platform\cite{OriginUnit}, created by Origin Quantum, facilitates the automated generation of quantum chip designs. Users can customize the layouts based on process specifications and automate the routing process. Current QEDA tools have made progress in their respective areas. However, they have not completely met the needs of users for thorough quantum chip design processes. The full potential of QEDA tools has not been fully realized.

The objective of this study is to develop a QEDA framework, referred to as EDA-Q, that has comprehensive design capabilities for quantum chips. The framework incorporates the fundamental features of existing QEDA tools, while also supplementing and optimizing the stages that are insufficiently covered by existing EDA tools. See Table \Ref{The Comparision of Different EDA Tools for Quantum Chip} for a detailed comparison of the EDA-Q framework with other quantum EDA tools. In addition, the framework is equipped with the algorithm library, chip component library, calculation library, process library, and simulation library to support the entire design process. EDA-Q adopts the architecture pattern of "Entity-Control-Process-Library Separation", providing the system with remarkable scalability and flexibility. EDA-Q's controller module employs a design approach called the "Generalized Functional Modules Model", ensuring compatibility between algorithms within the library and various chip components. Additionally, EDA-Q integrates numerous batch processing interfaces and utilizes a design scheme called the "Request Aggregation Mapping Model", offering users concise and flexible interactive interfaces.

The paper's organizational structure is outlined below. In Chapter 2, we present the complete design process of EDA-Q, including the beneficial features of each design stage. Chapter 3 provides a detailed explanation of the software architecture of EDA-Q, including the module interaction logic and the data transmission mechanism. It emphasizes the notable benefits of EDA-Q within this software architecture. Chapter 4 showcases the application of EDA-Q in real-world chip design, illustrating its beneficial impact on reducing design cycles, improving automation, and decreasing manual operation time. Chapter 5 presents a demonstration of the real efficiency of EDA-Q framework-designed chips, confirming their effectiveness in real-world chip design. Chapter 6 concludes by presenting an overview of the EDA-Q framework, highlighting its potential for expansion and providing useful references for future research.

\section{Architecture}
\begin{figure}
	\centering
	\includegraphics[width=\linewidth]{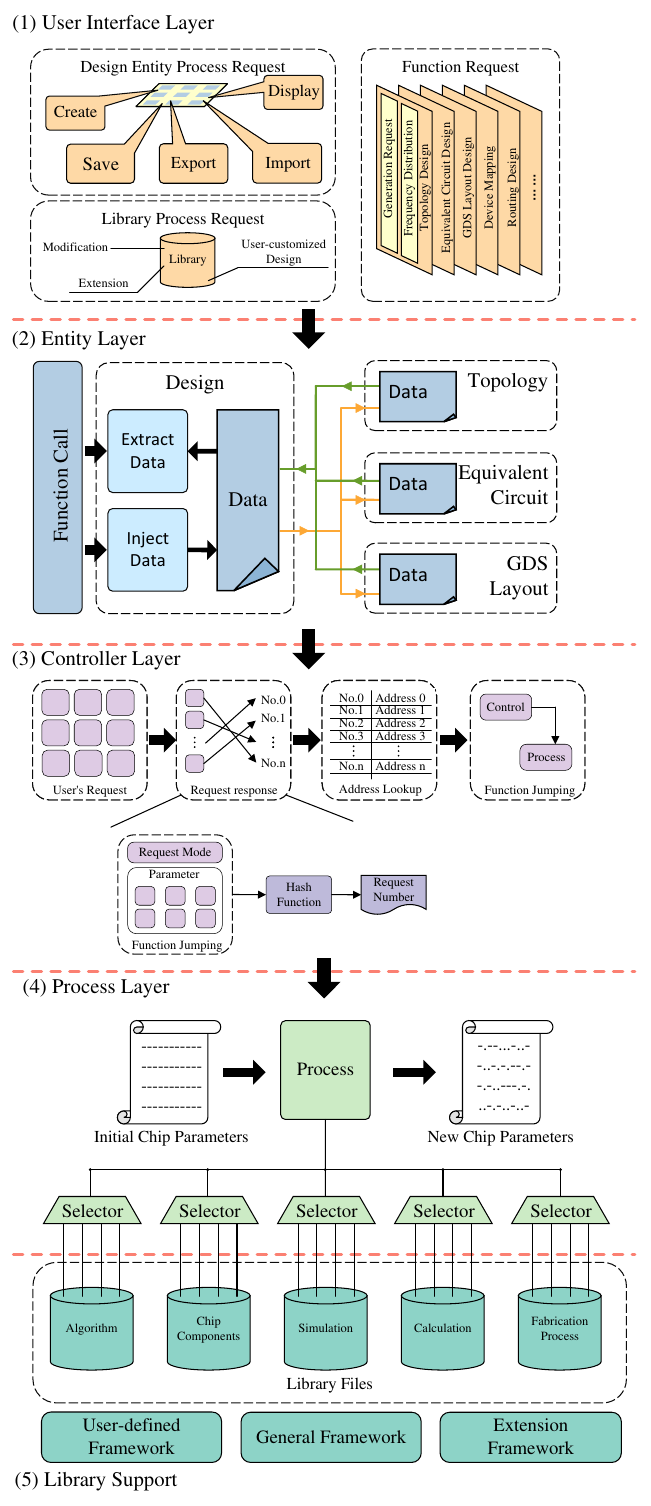}
	\caption{The Software Architecture of EDA-Q}
	\label{architecture}
\end{figure}

The comprehensive architecture of EDA-Q system is illustrated in Figure \ref{architecture}. This system embraces a structured software architecture model, systematically organized into five distinct layers: the User Interface Layer, the Entity Layer, the Controller Layer, the Process Layer, and the Library Support Layer. Each layer is designed to fulfill specific roles and responsibilities within the architecture, ensuring a robust, scalable, and efficient framework.
	
\paragraph{User Interface Layer}
The user interface layer serves as the intermediary between the user and the system, offering all the necessary functional interfaces for the user to design the chip using EDA-Q. The user's utilization of EDA-Q's functional modules can be considered as requests, and the requests can be categorized as Design Entity Process Request, Function Request, and Library Process Request. A Design Entity Process Request covers the operations of importing, exporting, and saving design files. A Function Request relates to the design of the current chip, which includes a range of operations in the design workflow discussed in the preceding chapter. A Library Process Request entails the expansion, customization, and modification of the system's support library. The User Interface Layer exclusively presents the interface that is visible to users of EDA-Q. User experience research and optimization are also conducted at this level. The complex logic involved in handling users' requests is executed for the subsequent level.
\paragraph{Entity Layer}
The Entity Layer consists of entities that carry all the information related to a single chip design, with the data within each entity uniquely identifying the design. Operations such as the export, import, and saving of design files essentially involve the manipulation of this entity data. Currently, in EDA-Q, there is an administrative entity known as the Design Entity, which is responsible for handling interactions with users and centrally managing other sub-entities. When a user requests parameters from the Design Entity, it employs a data extract method to extract data from the sub-entities and consolidate this information for delivery to the user. Upon receiving a request to update parameters, the Design Entity distributes the design parameters to various sub-entities for individual updates. At present, EDA-Q's sub-entities include the Topology Entity, the Equivalent Circuit Entity, and the GDS Layout Entity, each possessing data on qubit topology, equivalent circuits, and GDS layouts, respectively. These sub-entities operate independently but can also coordinate under the unified management of the Design Entity.
\paragraph{Controller Layer}
The Controller Layer categorizes user requests and arranges the algorithms to be employed based on the parameters contained in these requests. Based on experience with EDA-Q, it has been observed that it deals with a range of request types and intricate parameters during its execution. Various types of requests and their corresponding parameters are associated with distinct processing modules, resulting in intricate branching decisions. Thus, the Controller Layer utilizes a "request combination mapping" strategy to manage user requests. The figure illustrates the process of assigning a request number to each request based on its type and parameters. Next, it obtains the address of the processing function linked to the request number and redirects the program's execution to this address, entering the Process Layer. This approach efficiently resolves challenges associated with intricate control logic and challenges in maintaining and scaling the system. This method effectively addresses issues related to complex control logic and difficulties in maintenance and scalability, supporting EDA-Q in becoming a large-scale, general framework that facilitates future expansion of various functionalities and integration of state-of-the-art algorithms.
\paragraph{Process layer}
The Process Layer is accountable for carrying out certain functions within EDA-Q. Once the Controller Layer has determined the appropriate processing function using a sequence of control logic, the Process Layer obtains the required resources from the support library. Subsequently, it incorporates these resources into its own processing logic in order to successfully execute the assigned task. EDA-Q employs a parameterized methodology for processing, in which the design entities' parameters serve as the objects of the operations performed by the Process Layer. Modifications to the design are synonymous with changes in the design data. The updated data is subsequently distributed to the corresponding entities via the Entity Layer.
\paragraph{Support Library Layer}
The Support library Layer comprises a range of libraries that support the functioning of the system and provide specifications for the development of libraries. The support library comprises algorithm library, device library, simulation library, computation library, and process library. In EDA-Q, the execution of a function may necessitate the use of one or more algorithms. The algorithm resources required by the processing layer are organized within the algorithm library. Separating algorithms from processing logic enables easier expansion and maintenance of both. The device library offers a range of devices that come with equivalent circuits and GDS layouts. These resources are utilized for user designs. The simulation library contains the simulation modules of EDA-Q. The simulation process makes use of Python interfaces and methods from industrial software like Ansys and Comsol. The simulation system of EDA-Q is created by organizing and configuring these interfaces and methods, resulting in a unified interface style and concise calling logic. The calculation library incorporates mathematical algorithms for calculating the electrical characteristics of devices. The fabrication process library offers a framework for customizing processes, allowing the integration of processing specifications from different manufacturers during the manufacture of chips. It also includes a collection of universal process specifications. Alongside specific libraries, the support library layer encompasses library design specifications that outline the framework for users to modify and extend libraries.
\subsection{Architectural Advantages}
Quantum computing is currently in its early stages, with numerous research opportunities available in each step of chip design.  In this context, EDA-Q utilizes a separation architecture called "entity-control-process-library", which allows the system to have impressive scalability and flexibility.  Modifications to the approach of user requests can be made by adjusting the mapping rules in the control layer. System upgrades can be implemented by extending processing functions in the processing layer and adding content to the supporting library, while keeping the code kernel unchanged.

The design entity of EDA-Q and its sub-entities exclusively possess static storage capabilities, as well as methods for data extraction, access, updating, and display. The entities do not directly handle specific user requests. Instead, the control module processes all user requests. It extracts parameters from the entities and transfers them into the parameter processing workflow, which then updates the entities. The design pattern in question is called as the "extract-process-inject" pattern. The segregation of entities and functional modules in this design pattern enables the scalability of functional modules and fosters a coherent and sustainable organizational framework for the entities.

The device library of EDA-Q integrates a variety of chip devices, which often serve as the target objects for the same processing function. Traditional development approaches require the repetitive creation of identical processing functions for different devices. Moreover, when users add new devices to the library, these new devices are not compatible with the existing processing functions, leading to increased development costs and limited system scalability. To address this issue, EDA-Q employs a generic functional module in the development of its processing functions. The implementation of generic functional module involves a series of steps, which are not the focus of this paper. The emphasis is on how the design of this generic functional module significantly enhances development efficiency and system scalability.

\subsection{Data Flow}
When the user wants to perform a certain operation, he sends a request to the master control module through the api provided by the system. The control module organizes the control logic according to the request content, and usually sends instructions to the design entity to extract the chip design parameters. The whole design parameters may be extracted according to different user requests. It is also possible to extract some parameters of topology, equivalent circuit, GDS layout or other sub-entities, and then the control module organizes the response function module to perform parameter processing on the extracted parameters.   Different function modules will be organized to complete the parameter processing according to the different needs of users. After parameter processing, a new parameter will be returned to the design entity. The design entity updates each child entity according to the new parameters, and this time the user's request is processed.

\section{Design Workflow}
This chapter presents the design process of EDA-Q. The process can be categorized into topology design, equivalent circuit design, GDS layout design, device mapping, routing design, fabrication process mapping and simulation, based on a macroscopic level of granularity. Simultaneously, in practical design process, users may have the option to utilize specific stages or employ backtrack optimization based on the design context, depending on different application scenarios. EDA-Q also enables users to flexibly organize their own design schemes in response to these requirements.

\subsection{Topology Design}
The topological structure of qubits is crucial in multiple aspects of quantum chip, such as quantum error correction\cite{Zhao2021RealizationOA}, fault-tolerant fabrication\cite{PhysRevA.93.032322}\cite{Barends2014SuperconductingQC}, algorithm performance\cite{Liu2023}\cite{Hu2021PerformanceOS}, and so on. The stage determines the topological arrangement of qubits on the chip, defining the precise coordinates and the interconnections between qubits. Furthermore, incorporate topology-based optimization algorithms to enhance the physical implementation of qubits by optimizing their connectivity. Additionally, electrical parameter design that is related to the chip's topology can be performed, such as integrating a frequency allocation algorithm for qubits. EDA-Q provides two options for topology design: user-defined mode and circuit-defined mode. The user-defined mode enables users to create a customized topology from scratch. The circuit-defined mode employs a tailored topology to create a specialized quantum chip. By integrating topology mapping algorithms, the depth of the quantum line mapped to the topology can be reduced, and a topology specifically designed for a particular quantum algorithm can be accessed.

\subsection{Equivalent circuit design}
This module generates a visual equivalent circuit of circuit components based on the topological structure of the quantum chip and user-defined electrical parameters, aiding users in designing and calculating the corresponding Hamiltonian. It leverages calculation formulas from the computational library and equivalent circuit components from the component library to provide users with key parameter values for the quantum chip (such as \( E_C \), \( E_J \), etc.). These calculated values are visualized to assist users in analyzing and understanding quantum circuits. The EDA-Q interface allows for modifications to the equivalent circuits, enabling users to define design constraints based on specific needs, such as selecting different coupling modes, and rebuild the equivalent circuit accordingly.

Superconducting qubits can be modeled as anharmonic oscillators with non-uniform energy levels. They consist of a parallel combination of a capacitor and a Josephson junction, as illustrated in \( Q_1 \) in the figure below. The lowest two energy levels form the computational subspace of the qubit.

\begin{figure}[htbp]
    \centerline{\includegraphics[width=0.25\textwidth]{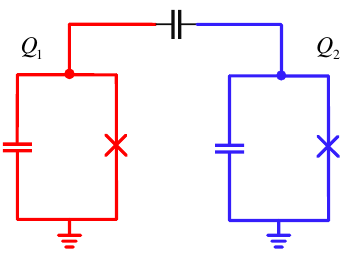}}
    \caption{The equivalent circuit of two superconducting qubits directly coupled by capacitance.}
    \label{device mapping}
  \end{figure}

In the design of large-scale superconducting quantum chips, it is essential to ensure that adjacent qubits have different frequencies to enable addressing of individual qubits, whether they are fixed-frequency qubits or tunable-frequency qubits. Additionally, the design of coupling strength determines the speed and fidelity of two-qubit gates.

In this work, we calculated the electrical parameters required for large-scale chip design, including qubit capacitance, Josephson junction resistance, and coupling capacitance between qubits. According to the following formulas:
\begin{equation}
{\omega _q} = {1 \mathord{\left/
 {\vphantom {1 {\sqrt {{L_j}{C_q}} }}} \right.
 \kern-\nulldelimiterspace} {\sqrt {{L_j}{C_q}} }} - {E_c}\label{eq}
\end{equation}
\begin{equation}
 {I_c}{R_n} = \frac{{\pi \Delta }}{{2e}}\tanh \left( {\frac{\Delta }{{2{k_B}T}}} \right)
\end{equation}
\begin{equation}
 {E_c} = \frac{{{e^2}}}{{2{C_q}}}
\end{equation}

Users only need to input the target qubit frequency and the target coupling strength, and we will calculate all the electrical parameters accordingly. Based on these calculated parameters, we further invoke the simulation module to determine the dimensions of each component. By using Maxwell simulation software, the capacitance matrix can be obtained, and the corresponding target capacitance values can then be derived through iterative optimization, as shown in the figure below.

\begin{figure*}[htbp]
    \centerline{\includegraphics[width=0.85\textwidth]{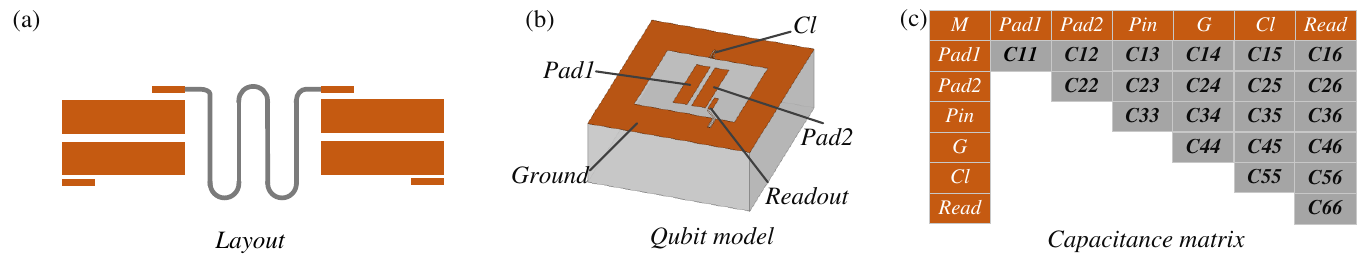}}
    \caption{Calculation of the capacitance matrix for the superconducting qubit model.}
    \label{fig:equivalent_circuit}
  \end{figure*}

\subsection{Chip Layout Design}
Chip Layout Design involves the precise arrangement and optimization of devices on a quantum chip, which is a graphical representation of the design system. The process consists of two distinct phases: layout generation and layout optimization. EDA-Q provides both automated and manual methods for generating chip layouts. The manual method enables users to arrange the chip layout in a flexible manner, supporting modular design by utilizing a chip component library to minimize the need for manual labor. The automated approach produces layouts using the pre-established topology, allowing users to incorporate design constraints (such as qubit types and distance) that are useful for the design on a large scale. EDA-Q also offers both manual and automated methods for optimizing chip layouts. Users can utilize batch processing functions or layout optimization algorithms to reduce mannual effort. The scalability of the algorithm library enables the incorporation of new algorithmic research and the application of various optimization strategies depending on the specific scenario and optimization target. For instance, it can be used for indium pillar layout optimization or for optimizing readout and coupling cavity layouts.

\subsection{Device mapping}
Designing the quantum devices with specific physical properties is a complex work that usually requires several cycles of design processes, precise manufacturing, and ongoing optimization procedures. Therefore, to ensure that the physical parameters of quantum devices have reached a relatively ideal state before actual chip fabrication, it is often necessary to conduct simulation verification and corresponding optimization. However, the simulation process is often time-consuming, with each simulation lasting for hours or even days, and each iteration necessitates human intervention. To solve these problems, EDA-Q has developed a device mapping module that generates devices with specific electrical characteristics according to user-defined indicators. The device mapper uses an automatic iterative optimization algorithm, along with the assistance of calculation library, simulation library and chip component library, to achieve the automatic design of the entire component. The procedure is illustrated in figure \ref{device mapping}.

\begin{figure}[h]
	\centering
	\includegraphics[width=\linewidth]{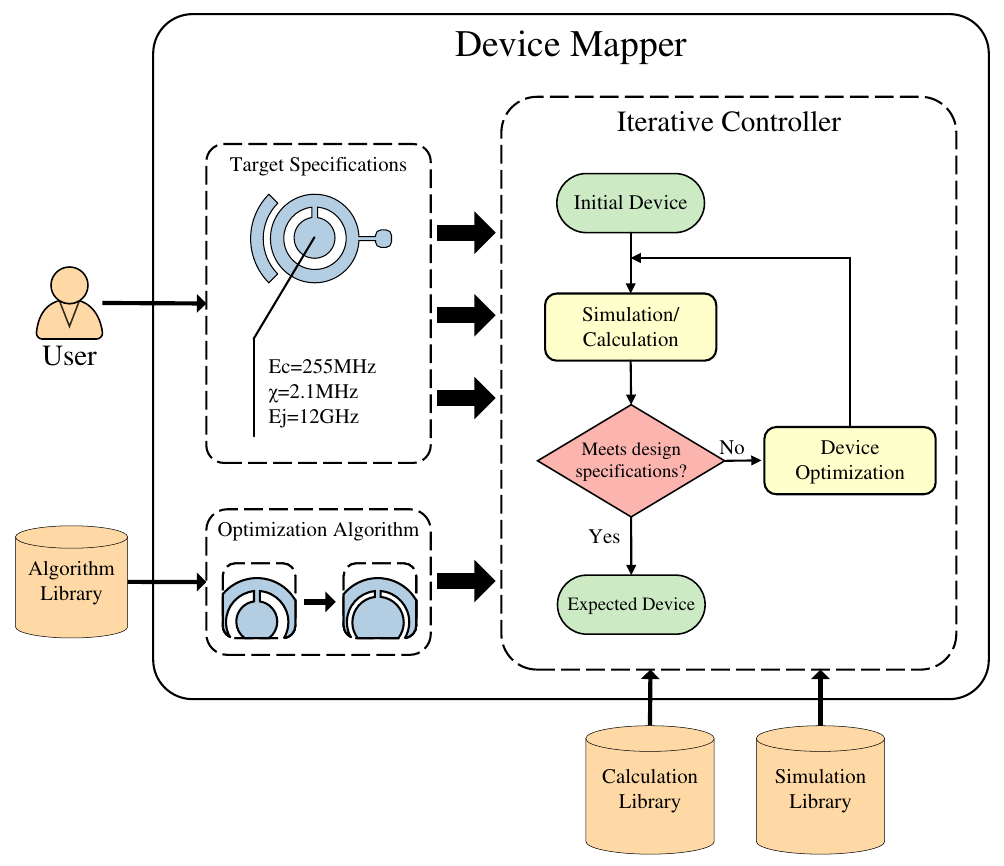}
	\caption{Working architecture of device mapper}

	\label{device mapping}
\end{figure}

\subsection{Routing design}
The Routing Design process entails the strategic positioning of transmission lines, control lines, and other necessary wirings to accommodate the current chip layout.   EDA-Q offers three routing modes: global automatic routing, semi-automatic routing, and manual routing. The manual routing mode allows users to completely customize the path of the wirings for precise routing, making it ideal for small-scale and highly customized chip design.  The global automatic routing mode utilizes the routing algorithm from the routing algorithm library to perform automatic routing in various scenarios, taking into account the current GDS layout. Semi-automatic routing is a method that combines user instructions with routing algorithms to make customized routing design easier. Due to its combination of precision and automation, this approach is widely employed during the optimization phase of routing design process. 

In the physical design of superconducting quantum chips, the premise of routing is the realization of the layout. Here, we uniformly adopt a matrix dot-style qubit layout as the foundation for our routing initiatives. At this point, our EDA-Q has already accomplished the generation of logical topologies and the construction of actual qubits. Building upon this foundation, SQCR presents two distinct routing strategies: SQCR-Maze and SQCR-Pattern. After routing, components such as air bridges and indium columns are added to form a complete layout.

Classical routing algorithms like Lee's Maze algorithm, employing methods such as BFS and Dijkstra's, are foundational in the search for efficient paths under congestion \cite{b40}. The A* algorithm, with its heuristic approach, further refines the process by balancing actual and estimated costs, offering significant advantages in practical routing scenarios \cite{b41}, \cite{b42}, \cite{b43}. Applied to superconducting quantum chips, the SQCR-Maze method adapts A* to meet the unique demands of minimizing line length and avoiding crosstalk in quantum circuits, thereby enhancing signal integrity and reducing noise interference \cite{b44}.

\begin{figure}[htbp]
  \centerline{\includegraphics[width=0.45\textwidth]{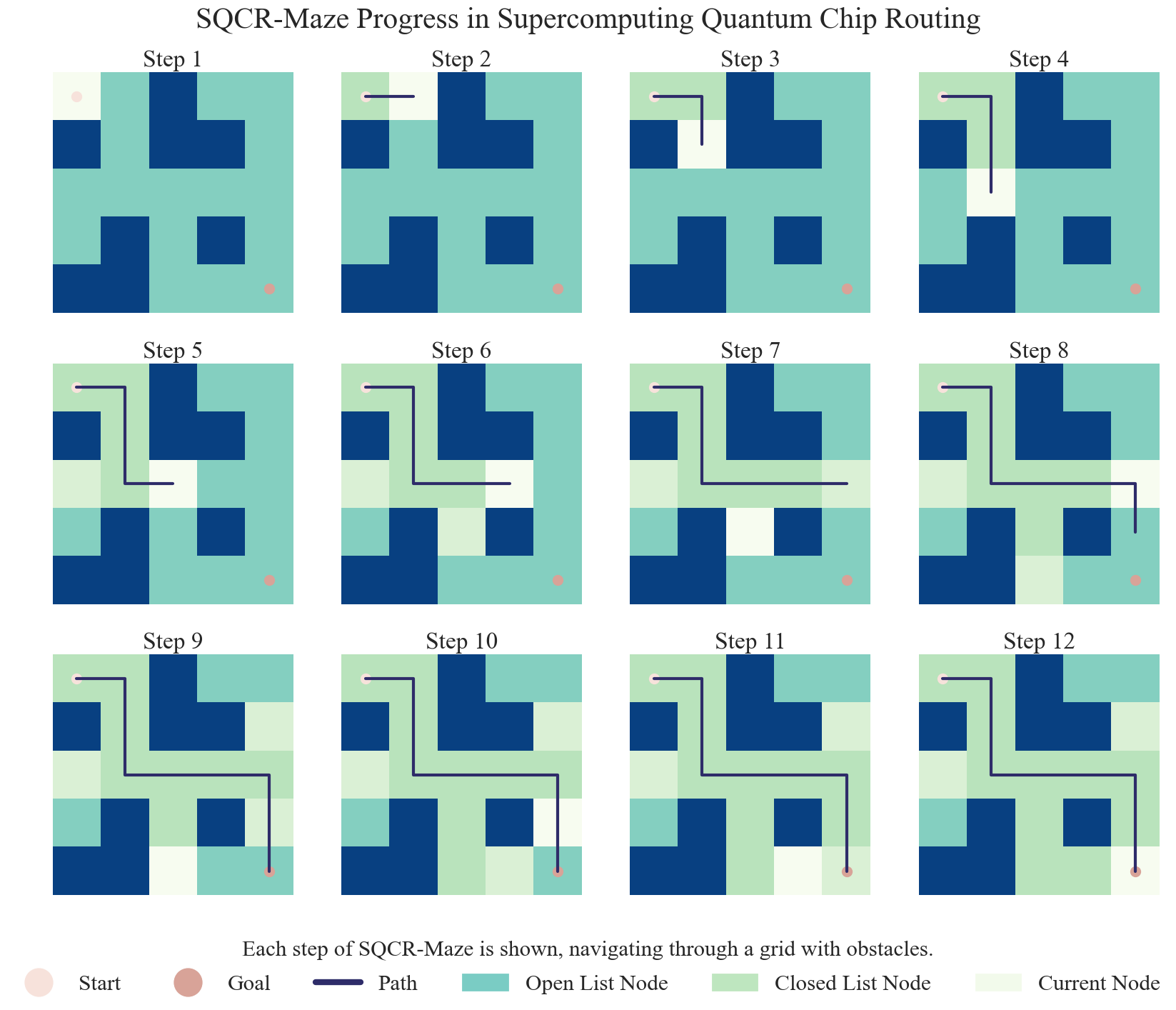}}
  \caption{The Process of SQCR-Maze Pathfinding Algorithm}
  \label{fig}
\end{figure}

The key to SQCR-Maze lies in the design of the heuristic function. The heuristic function \( h(n) \) evaluates the estimated cost from the current node \( n \) to the destination. In the absence of direct information, geometric distance is commonly used for heuristic evaluation. If only vertical or horizontal movements are allowed in the grid, the Manhattan distance is typically used: \( h(n) = |x_n - x_g| + |y_n - y_g| \). If diagonal movements are permitted, the Euclidean distance can be used: \( h(n) = \sqrt{(x_n - x_g)^2 + (y_n - y_g)^2} \). Here, \( (x_n, y_n) \) are the coordinates of the current node \( n \), and \( (x_g, y_g) \) are the coordinates of the destination.

To further reduce crossings(use of air bridges in single-layer chips) and turns(avoid exposure to qubis and resonators in single-layer chips), the heuristic function can be adjusted to increase the expected cost for corners and crossings:
\[ h(n) = d(n, g) + k \cdot (E_c + E_x) \]
Here, \( d(n, g) \) is the geometric distance (Manhattan or Euclidean) from node \( n \) to the goal \( g \), \( E_c \) is the estimated minimum number of corners from \( n \) to \( g \), \( E_x \) is the estimated minimum number of crossings, and \( k \) is a tuning factor to balance the impact of geometric distance against corners and crossings.

The number of corners \( E_c \) can be estimated by analyzing the planned path from the current node to the target node. If each step of the path moves from one grid cell to another, \( E_c \) can be calculated by comparing the direction of two consecutive steps on the path. If the direction changes, it indicates a corner on the path. \( E_c \) can be represented as: \( E_c = \sum_{i=2}^{n-1} \delta(d_{i-1}, d_i) \). Here, \( n \) is the total number of nodes on the path, \( d_i \) is the direction from node \( i-1 \) to node \( i \), and \( \delta(d_{i-1}, d_i) \) is a function that returns 1 when \( d_{i-1} \neq d_i \) and 0 otherwise.

To calculate the number of crossings \(E_x\) in grid-based path planning, an occupancy matrix \(M\) is employed. This matrix corresponds in size to the planning area, initially set with all entries \(M_{ij} = 0\) to denote unoccupied cells. As paths are generated, each step checks for crossings by examining if the cell at each point \(p_k\) in the path sequence \(P = \{p_1, p_2, \dots, p_n\}\) is already marked. If \(M_{p_k} = 1\), indicating an existing mark, a crossing is registered and \(E_x\) is incremented. Thus, the total number of crossings is calculated as \(E_x = \sum_{k=1}^{n} I(M_{p_k} = 1)\), where \(I\) is the indicator function that identifies a crossing \cite{b29}.

In SQCR-Maze, the start node is first added to the open list. Subsequently, the algorithm repeatedly selects the node with the lowest cost from the open list as the current node and moves it to the closed list. At the same time, it considers all its reachable and not yet closed neighbor nodes, adding them to the open list or updating their costs in the open list. This process continues until the target node is found or the open list is empty. In this way, the algorithm gradually constructs the optimal path from the start point to the target point, ensuring that nodes that have already been considered are not revisited. Figure 1 shows the pathfinding process applied in a superconducting quantum chip using SQCR-Maze.

The application of SQCR-Maze sometimes results in delays, especially when creating direct point-to-point connections, which are prevalent in the structured and replicable component layouts of many superconducting quantum chips. In multi-dimensional quantum chip architectures, the extensive routing space substantially reduces the effects of crossing lines. Routinely, these circuits are designed to be succinct with minimal bends. This paper proposes a standardized routing method named SQCR-Pattern, which utilizes predefined routing patterns to achieve efficient and error-free circuit layouts. SQCR-Pattern ensures cross-free connections and meets device connectivity requirements by employing a predefined global routing strategy. The methodology also includes the optimization of routing dimensions, such as the width and spacing of lines, to maintain the physical viability of patterned routings\cite{b46,b47,b48,b49}.

  \begin{figure*}[htbp]
    \centerline{\includegraphics[width=0.7\textwidth]{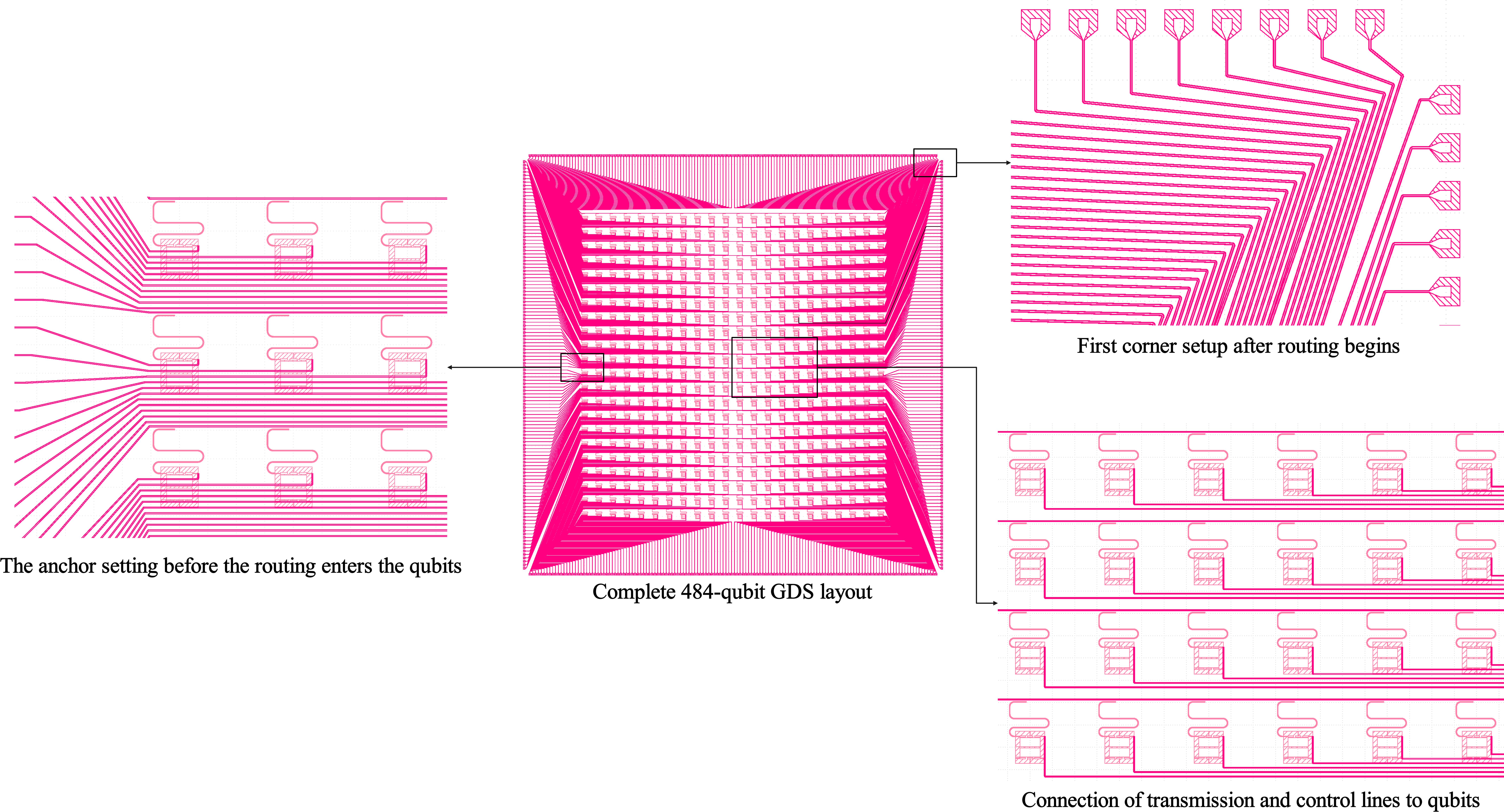}}
    \caption{SQCR-Pattern Results for a 484-qubit Chip}
    \label{fig}
  \end{figure*}

The initial step in the SQCR-Pattern method is pin allocation, which is crucial for connecting and controlling qubits based on several parameters such as the topology of qubits, their positions, readout routings configurations, and chip dimensions. Pins serve as interfaces for controlling and reading signals from qubits. Initially, basic attributes of pins, such as distance to the chip, pad width, and pad gap, are set based on user process requirements. The pin count is then calculated for each direction (top, bottom, left, right) based on the dimensions of the qubits and the chip. This calculation determines the minimum chip length required for the pins in each direction, and the chip size is adjusted if it is insufficient to accommodate all pins. The final step involves setting the start and end positions of the chip, with the parity of the columns playing a critical role in determining the rounds of pin allocation. The calculation of pin quantity in the SQCR-Pattern example is a fundamental task that precedes the routing process.

In optimizing pin distribution on superconducting quantum chips, calculate the total number of pins based on the \(m \times n\) grid layout as \( \text{Total Pins} = 2m + mn \), where \(2m\) represents pins per row for transmission lines and \(mn\) for qubit connections. Distribute pins evenly across the chip's edges by assigning rows to top, bottom, and sides, balancing pin count to minimize discrepancies across edges. This strategic allocation ensures efficient layout and connectivity, critical for optimal chip performance.

% \begin{figure}[htbp]
%     \centerline{\includegraphics[width=0.45\textwidth]{3.png}}
%     \caption{Implementation process of SQCR-Pattern method}
%     \label{fig}
% \end{figure}

After completing the pin assignment, the SQCR-Pattern method proceeds with the routing. For superconducting quantum chips using flip-chip assembly, control and transmission lines can be routed on the opposite layer corresponding to the qubit area. This method avoids crossing with coupling cavities.

In the SQCR-Pattern for quantum chips, routing strategies begin at the chip's periphery and extend to qubit or readout routings. This approach employs a bilaterally optimized path to minimize interference. Control lines are carefully routed to avoid crossing with transmission lines, and control pins at various positions are mapped strategically to correspond with qubits. For top and bottom pins, the mapping is either in reverse order (\(M[S[i]] = Q[|S|-i-1]\)) or in forward order (\(M[S[i]] = Q[i]\)), depending on their positions. For the side pins, an alternating strategy is used, with left side in forward and right side in reverse order. This methodical routing plan addresses the challenge of connecting multiple qubit systems with minimized crosstalk and optimized path efficiency. The systematic segment analysis and mapping provide a robust framework for routing in complex quantum computing architectures.
Figure 2 showcases the patterned routing results for a 484-qubit chip, demonstrating the sophisticated routing schema pivotal for advancing quantum computing architectures.

From the perspective of time complexity, the SQCR-Pattern has significant advantages. Before initiating the routing process, all qubits on the chip are generated, including the overall parameters of the chip and qubits. Based on this, we perform device routing on chips with known qubit scales. Our QEDA can directly generate layouts based on user-specified configurations without any manual intervention. Among existing QEDA tools, only SpinQ's Tianyi supports fully automated routing; other QEDA software requires manual intervention during the routing process, making it impossible to measure routing time. Tianyi's layout is fixed, representing a simpler target within the SQCR-Pattern. Currently, Tianyi supports up to 200 qubits, and we use it for comparison with SQCR. We tested the runtime of three methods on qubit scales ranging from 2x2 to 28x28 (Tianyi EDA supports up to 14x14 scale) on a computer equipped with a 13th Gen Intel(R) Core(TM) i9-13900KF 3.00 GHz processor, 64.0 GB (63.7 GB usable) memory, and Windows 11 Pro operating system, as shown in Figure 5. The application of SQCR-Maze causes the program runtime to increase quadratically with the scale of qubits. In contrast, the runtime of SQCR-Pattern grows linearly with the scale of qubits, comparable to the routing time of Tianyi EDA. This demonstrates that SQCR-Pattern not only has theoretical complexity advantages but can also benchmark against solutions integrated into existing software.

In terms of generality and scalability, SQCR-Maze undoubtedly has an advantage. Its heuristic search process can be applied to superconducting quantum chip routing of various sizes and architectures. The design of the algorithm relies on a high level of abstraction from the chip itself, primarily depending on the construction of grid graphs. In contrast, the freedom of SQCR-Pattern is greatly limited. Once a routing pattern is determined, the settings for corners and the allocation of routing resources are fixed, making them difficult to adjust. Any changes could significantly impact the entire routing strategy.

\begin{figure}[htbp]
    \centerline{\includegraphics[width=0.45\textwidth]{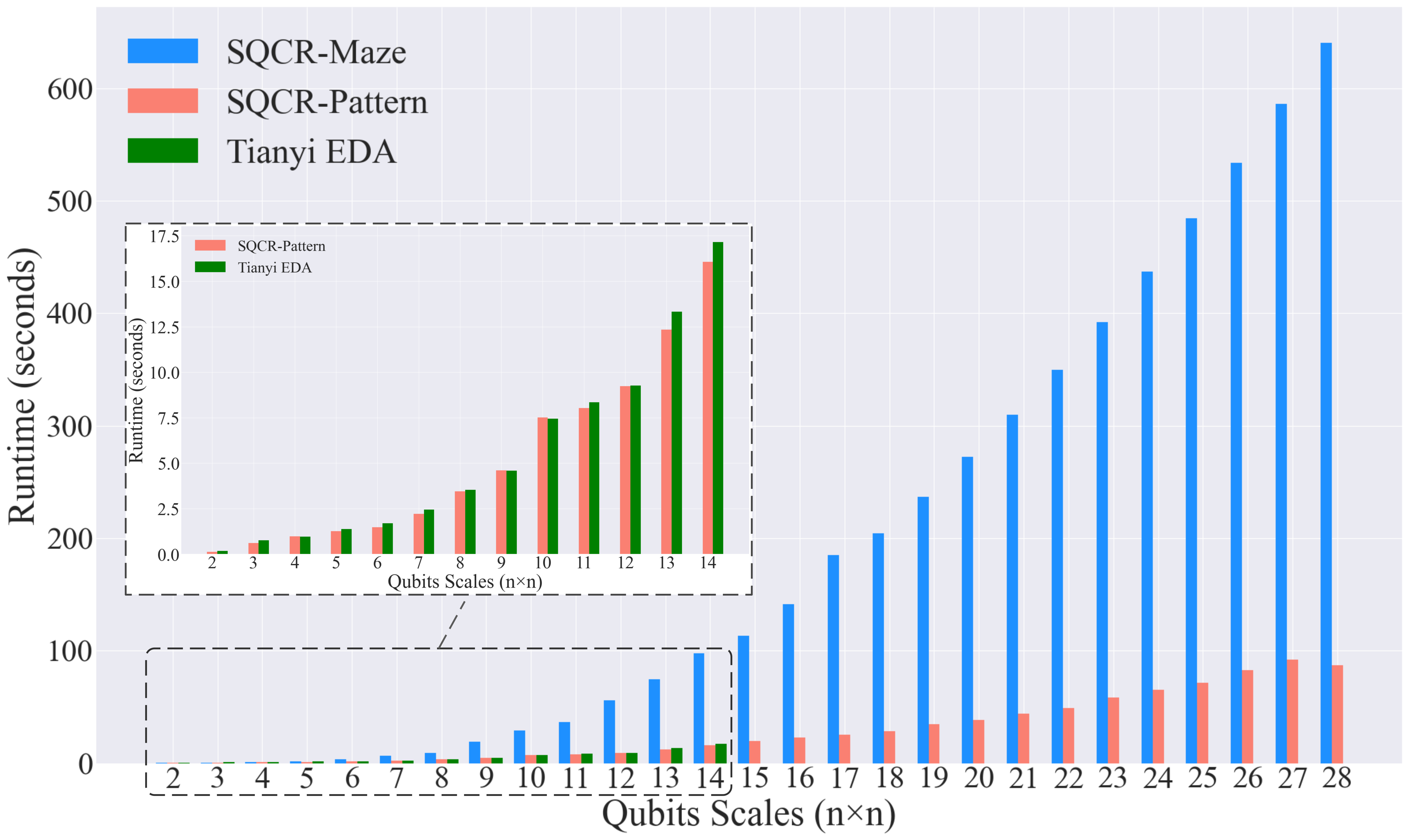}}
    \caption{Comparison of the runtime of SQCR and Tianyi EDA routing at different qubit scales}
    \label{fig}
\end{figure}

In summary, the two methods included in SQCR each have their own advantages and disadvantages in terms of scalability and time complexity, making them suitable for different application scenarios. Under conditions of multilayer architecture and clear processes, SQCR-Pattern has an edge. However, it struggles with transferability. On the other hand, for quantum chip designs that are complex and difficult to define with patterned rules, the use of SQCR-Maze, which has a higher time complexity, is recommended.

\subsection{Fabrication Process Mapping}
In this stage, the chip layout is modified to adhere to the process specification. The process mapper utilizes the process specifications stored in the fabrication process library to modify the dimensions and characteristics of devices, such as line width, corner radius, pin size, and air bridge. It also combines the cost or electrical loss associated with different process techniques in order to optimize a weighted optimal chip layout. The fabrication process library can integrate commonly used process specifications that are currently available in the market, while also providing a customizable interface for user-defined process standards.

\subsection{Simulation}
Industrial simulation software such as ANSYS and COMSOL is employed to validate the design through simulation. EDA-Q offers integrated interfaces in its simulation library for various simulation modes, enabling users to perform simulation analyses on chip devices through simple function calls and parameter passing. Furthermore, it can be paired with an integrated device optimizer to automate the optimization and adjustment process, thereby achieving the desired electrical performance.

\subsubsection{Qubit Parameter Simulation}

Leveraging three-dimensional finite element analysis methods, we obtained the dimensions of the quantum bits designed by users and constructed models and calculated parameters for planar X-mon, Flip-Chip X-mon, and planar Transmon qubits. The specific parameters are presented in Table~\ref{tab:parameters}.

\begin{figure*}[htbp]
    \centerline{\includegraphics[width=0.85\textwidth]{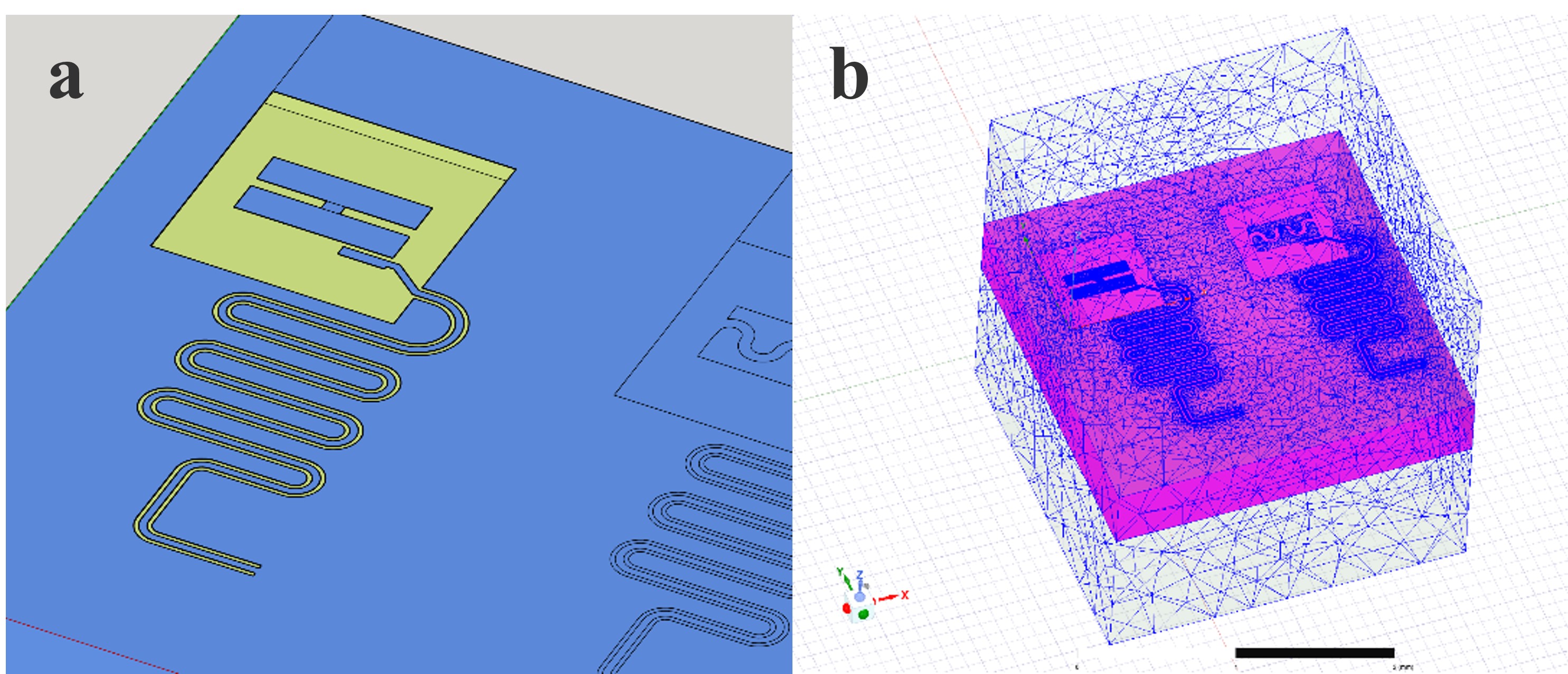}}
    \caption{Planar floating Transmon structure and three-dimensional finite element mesh partitioning: (a) Schematic diagram of superconducting qubit model. (b) Schematic diagram of grid division for superconducting quantum chip.}
    \label{fig:model}
\end{figure*}

The superconducting qubit models constructed using the QEDA tool are based on the Transmon structure, which includes Josephson junctions, suspended capacitors, and readout resonant cavities, among others. An example of such a model is shown in Figure \ref{fig:model}(a). During the simulation of superconducting quantum chips, the finite element analysis method is employed to mesh the chip grid, shown in Figure \ref{fig:model}(b).

\begin{table}[htbp]
    \centering
    \caption{\label{tab:parameters}Input and output parameters for Qubit modeling, including the qubit structure and electrical parameters of EDA-Q Simulation.}
    \begin{tabular}{ll}
        \toprule
        \textbf{Input} & \\
        \midrule
        Quantum Bit Size & \begin{tabular}[c]{@{}l@{}}Capacitor Shape\\ Capacitor Size\\ Josephson Junction Type\\ Planar Xmon\\ Flip-Chip Xmon\\ Planar Transmon\\ \dots\end{tabular} \\
        Coupling Pad Design & \begin{tabular}[c]{@{}l@{}}Number of Coupling Pads\\ Size of Coupling Pads\end{tabular} \\
        Capacitor Parameters & \begin{tabular}[c]{@{}l@{}}Quantum Bit Self Capacitance\\ Quantum Bit Coupling Capacitance\\ Parasitic Capacitance\\ Capacitance Energy\\ Quantum Bit Frequency\end{tabular} \\
        Josephson Junction Parameters & \begin{tabular}[c]{@{}l@{}}Josephson Energy\\ Josephson Equivalent Inductance Value\\ Room Temperature Resistance\\ Critical Current\end{tabular} \\
        Field Parameters & \begin{tabular}[c]{@{}l@{}}Surface Current Density\\ Electric Field Intensity Distribution\\ Magnetic Field Intensity Distribution\end{tabular} \\
        Other Parameters & \begin{tabular}[c]{@{}l@{}}Non-harmonicity\\ \dots\end{tabular} \\
        \midrule
        \textbf{Output} & \\
        \midrule
        Capacitor Parameters & \begin{tabular}[c]{@{}l@{}}Quantum Bit Self Capacitance\\ Quantum Bit Coupling Capacitance\\ Parasitic Capacitance\\ Capacitance Energy\\ Quantum Bit Frequency\end{tabular} \\
        Josephson Junction Parameters & \begin{tabular}[c]{@{}l@{}}Josephson Energy\\ Josephson Equivalent Inductance Value\\ Room Temperature Resistance\\ Critical Current\end{tabular} \\
        Field Parameters & \begin{tabular}[c]{@{}l@{}}Surface Current Density\\ Electric Field Intensity Distribution\\ Magnetic Field Intensity Distribution\end{tabular} \\
        Other Parameters & \begin{tabular}[c]{@{}l@{}}Non-harmonicity\\ \dots\end{tabular} \\
        \bottomrule
    \end{tabular}
\end{table}

% \begin{figure}
% \centering
% \includegraphics[width=1\linewidth]{model.jpg}
% \caption{\label{fig:model}Planar floating Transmon structure and three-dimensional finite element mesh partitioning: (a) Schematic diagram of superconducting qubit model. (b) Schematic diagram of grid division for superconducting quantum chip.}
% \end{figure}

In the simulation of superconducting quantum chips, some electrical parameters need to be derived through theoretical calculations. The capacitance matrix corresponding to the chip's GDS is obtained using three-dimensional finite element simulation methods \cite{b55}. The self-capacitance of the qubit is calculated through the coupling capacitance between the capacitive plate and the grounded part \cite{b54}. The self-capacitance $Cq$ of the X-mon is expressed as \cite{b52}:
\begin{equation}
C_q = -\text{float}\left(df_1.loc['bt\_Xmon', 'ground\_Q\_chip\_plane']\right)
\end{equation}

The charge energy \( E_C \) is given by:
\begin{equation}
 E_C = \frac{e^2}{4\pi \times 10^{21} \times C_q \times \hbar} 
\end{equation}

The Josephson energy \( E_J \) is given by:
\begin{equation}
 E_J = \left( \frac{f_q \times 10^9 + E_C}{8 \times E_C} \right)^2 
\end{equation}

The Josephson critical current \( I_c \) is:
\begin{equation}
I_c = \frac{E_J \times h}{\Phi_0}
\end{equation}

The normal resistance at room temperature \( R_n \) is:
\begin{equation}
R_n = \frac{\pi \times 0.182 \times 10^{-3}}{2 \times I_c}
\end{equation}

The equivalent Josephson inductance \( L_j \) is:
\begin{equation}
L_j = \frac{\hbar}{2eI_c \cos(\delta)}
\end{equation}

where \( \hbar \) is the reduced Planck constant with a value of \( 6.6260755(40) \times 10^{-34} \, \text{J} \cdot \text{s} \), \( I_c \) is the Josephson critical current, and \( \delta \) represents the phase difference across the Josephson junction.

\subsubsection{Resonator Parameter Simulation}
By acquiring values such as the CPW length \( L \) and type settings from the GDS layout, the width \( w \) of the center conductor and gap \( g \) of the coplanar waveguide, and the dielectric constant \( \epsilon \) of the substrate material for the superconducting quantum chip, a three-dimensional simulation model of the resonator section can be established. The three-dimensional model is meshed, and finite element simulation methods \cite{b55} are utilized to perform large-scale frequency sweeps and accurately approach the resonator frequency using both driven and eigenmode approaches. Additionally, the \( S_{21} \) curve and quality factor \( Q_c \) of the resonator can be obtained based on the calculation process, with specific parameters shown in Table~\ref{tab:resonator} \cite{b53}.

\begin{table}[htbp]
    \centering
    \caption{\label{tab:resonator}Input and Output Parameters for Coplanar Waveguide (CPW) and Resonator Modeling.}
    \begin{tabular}{ll}
        \toprule
        \textbf{Input} & \\
        \midrule
        CPW Dimensions & \begin{tabular}[c]{@{}l@{}}Resonator Length\\ $\lambda/2$ Resonator\\ $\lambda/4$ Resonator\\ Center Conductor Width\\ Gap Width\\ Coupling Length with Feedline\\ Metal Deposition Thickness\end{tabular} \\
        Substrate Parameters & \begin{tabular}[c]{@{}l@{}}Substrate Type\\ Substrate Dielectric Constant\\ Substrate Thickness\end{tabular} \\
        Other Parameters & \begin{tabular}[c]{@{}l@{}}Magnetic Penetration Depth\\ Center Conductor Current\end{tabular} \\
        \midrule
        \textbf{Output} & \\
        \midrule
        Resonator Parameters & \begin{tabular}[c]{@{}l@{}}Readout Resonator Frequency\\ External Quality Factor \( Q_l \)\\ Internal Quality Factor \( Q_i \)\\ Energy Dissipation Rate\\ Surface Current Density\\ Electric Field Intensity Distribution\\ Magnetic Field Intensity Distribution\\ Dispersion Shift\\ \dots\end{tabular} \\
        \bottomrule
    \end{tabular}
\end{table}

Employing three-dimensional finite element simulation analysis, the S21 curve of the coplanar waveguide resonator corresponding to the GDS layout was obtained, and the valley of the curve was extracted as the design frequency. The frequency scan process under eigenmode and the S21 diagram under driven mode are shown in Figure \ref{fig:S21}:

\begin{figure}[htbp]
    \centerline{\includegraphics[width=0.45\textwidth]{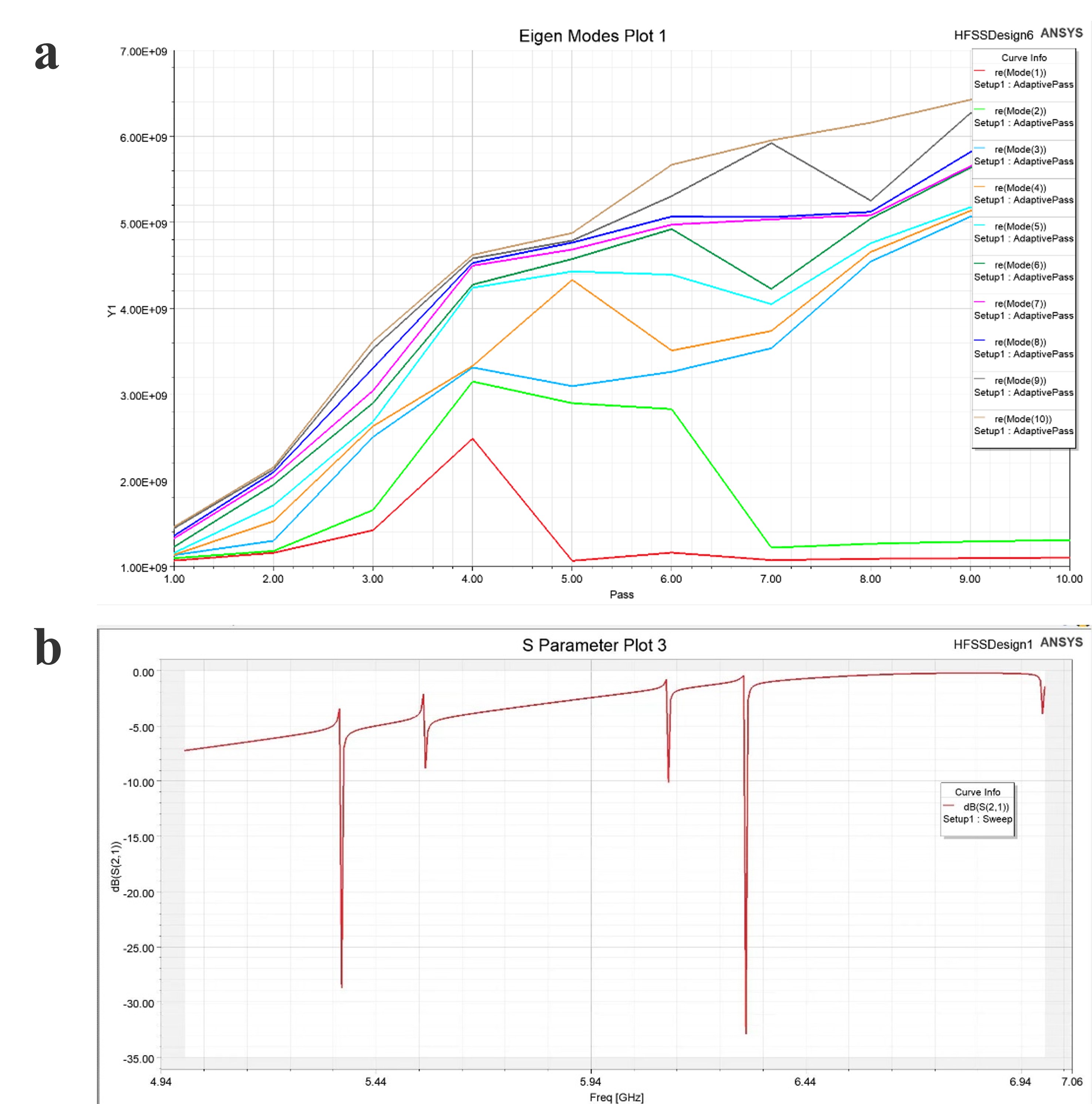}}
    \caption{Schematic diagram of resonant cavity frequency simulation results: (a) The eigenmode simulation yields the resonator frequency field solution. (b) The driven-mode simulation provides the resonator S21 curve.}
    \label{fig:S21}
\end{figure}
% \begin{figure}
% \centering
% \includegraphics[width=0.7\linewidth]{S21.jpg}
% \caption{\label{fig:S21}Schematic diagram of resonant cavity frequency simulation results: (a) The eigenmode simulation yields the resonator frequency field solution. (b) The driven-mode simulation provides the resonator S21 curve.}
% \end{figure}

\section{Application Example}
We used EDA-Q to create an entire 64-bit chip configuration. The Figure 12 displays the images corresponding to each design stage. The chip has undergone topology design, equivalent circuit design, chip layout design, automatic routing, and ultimately performed device mapping, process mapping, and adjustment of details to achieve the final GDS layout.

\begin{figure}
    \centering
    \begin{subfigure}[b]{0.45\linewidth}
        \includegraphics[width=\linewidth]{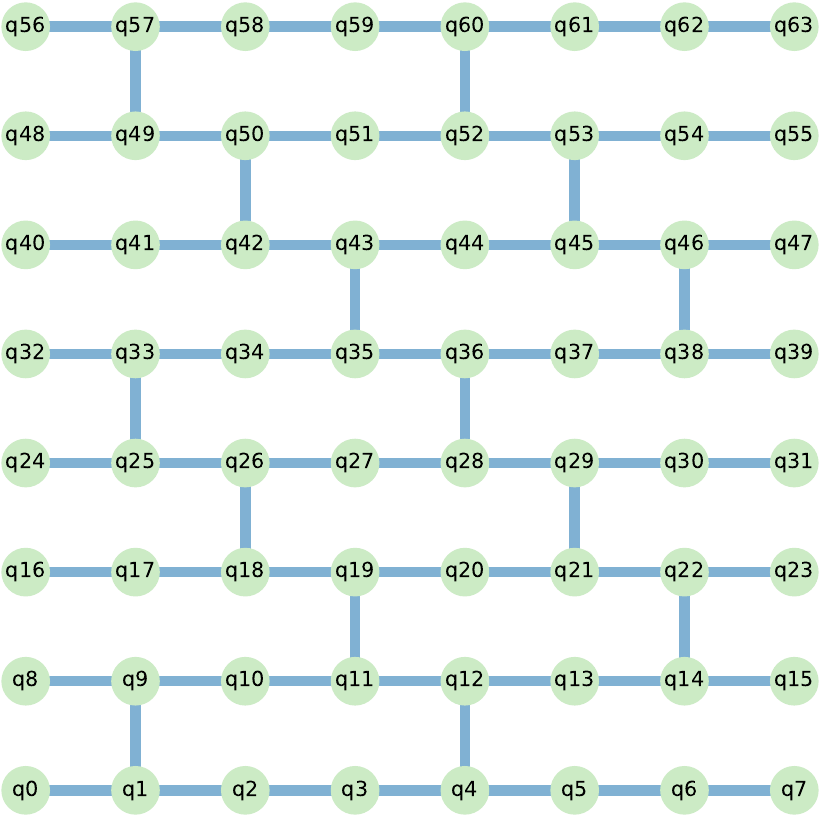}
        \caption{Topology Design}
    \end{subfigure}
    \hfill
    \begin{subfigure}[b]{0.45\linewidth}
        \includegraphics[width=\linewidth]{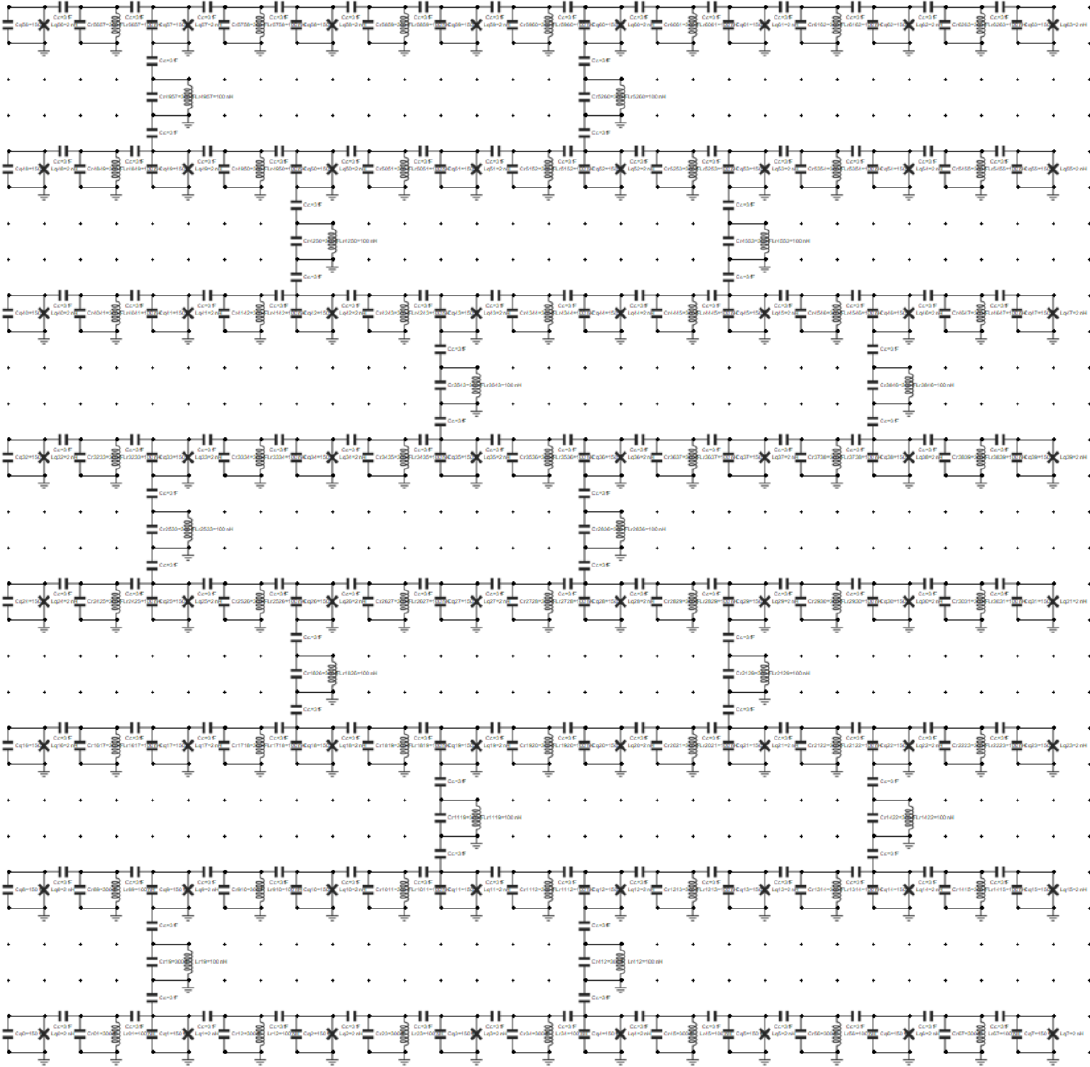}
        \caption{Equivalent Circuit Design}
    \end{subfigure}
    \begin{subfigure}[b]{0.45\linewidth}
        \includegraphics[width=\linewidth]{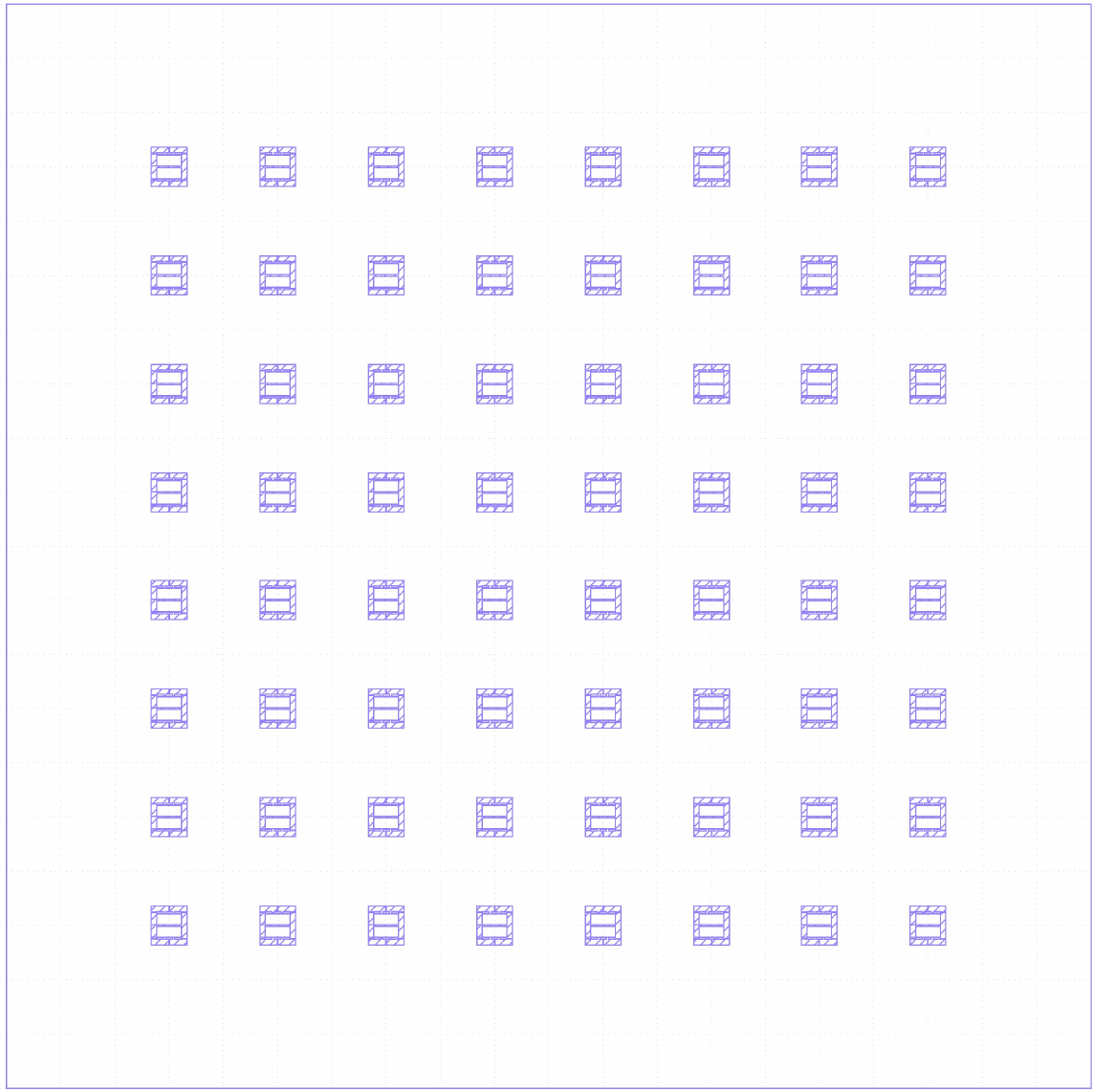}
        \caption{Qubits Layout}
    \end{subfigure}
    \hfill
    \begin{subfigure}[b]{0.45\linewidth}
        \includegraphics[width=\linewidth]{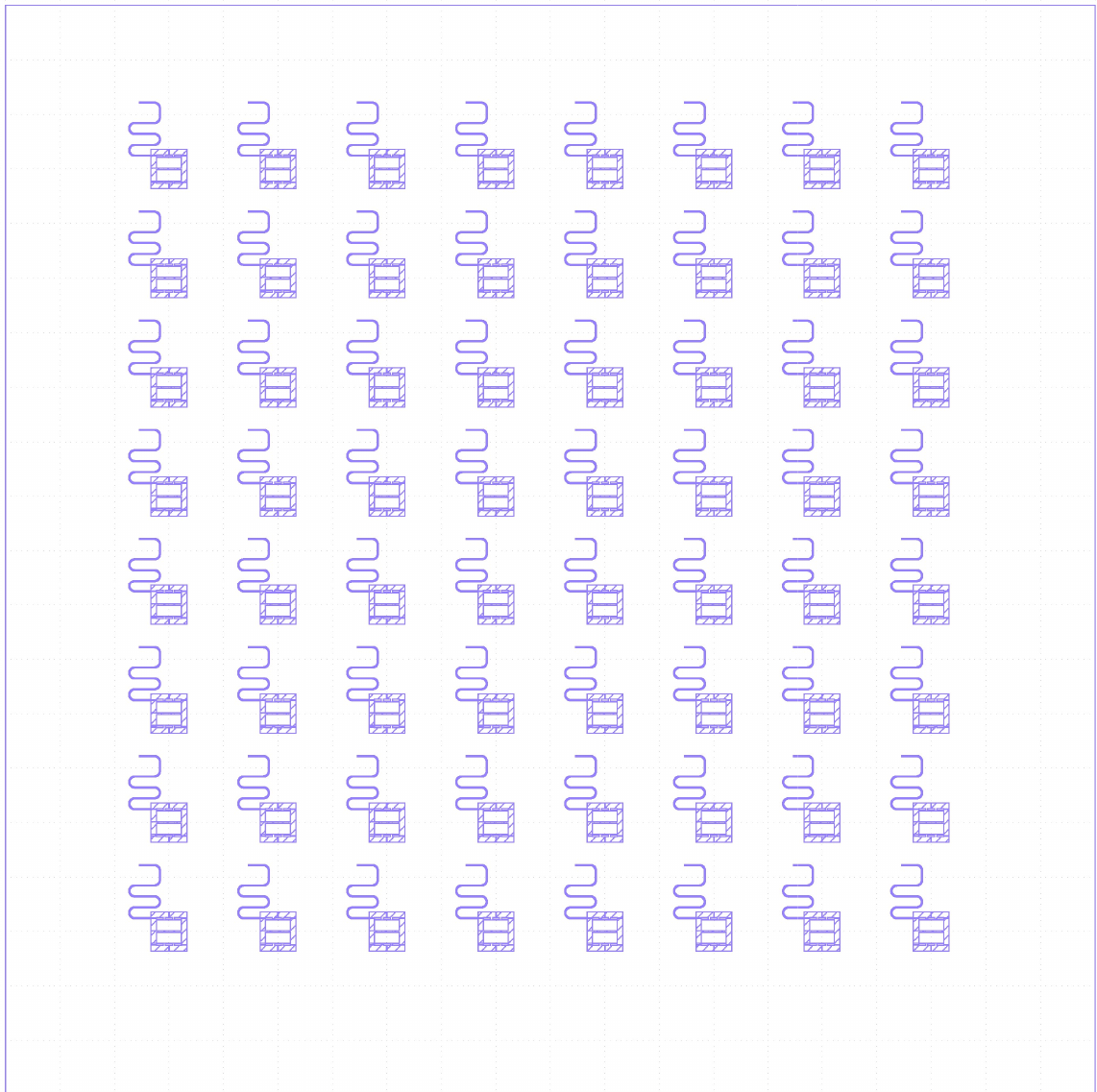}
        \caption{Readout Cavity Generation}
    \end{subfigure}
    \begin{subfigure}[b]{0.45\linewidth}
        \includegraphics[width=\linewidth]{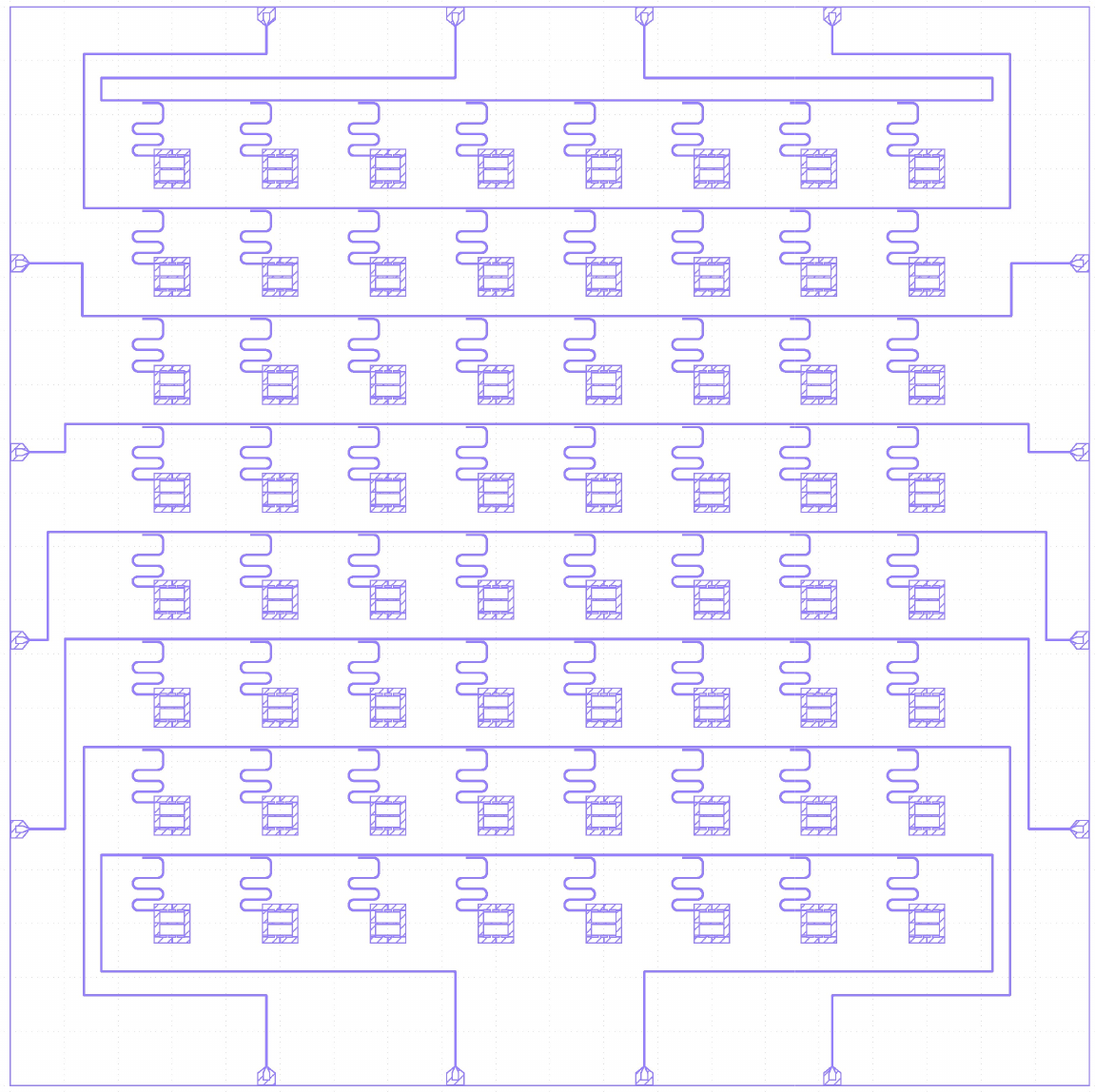}
        \caption{Global Automatic Routing}
    \end{subfigure}
    \hfill
    \begin{subfigure}[b]{0.45\linewidth}
        \includegraphics[width=\linewidth]{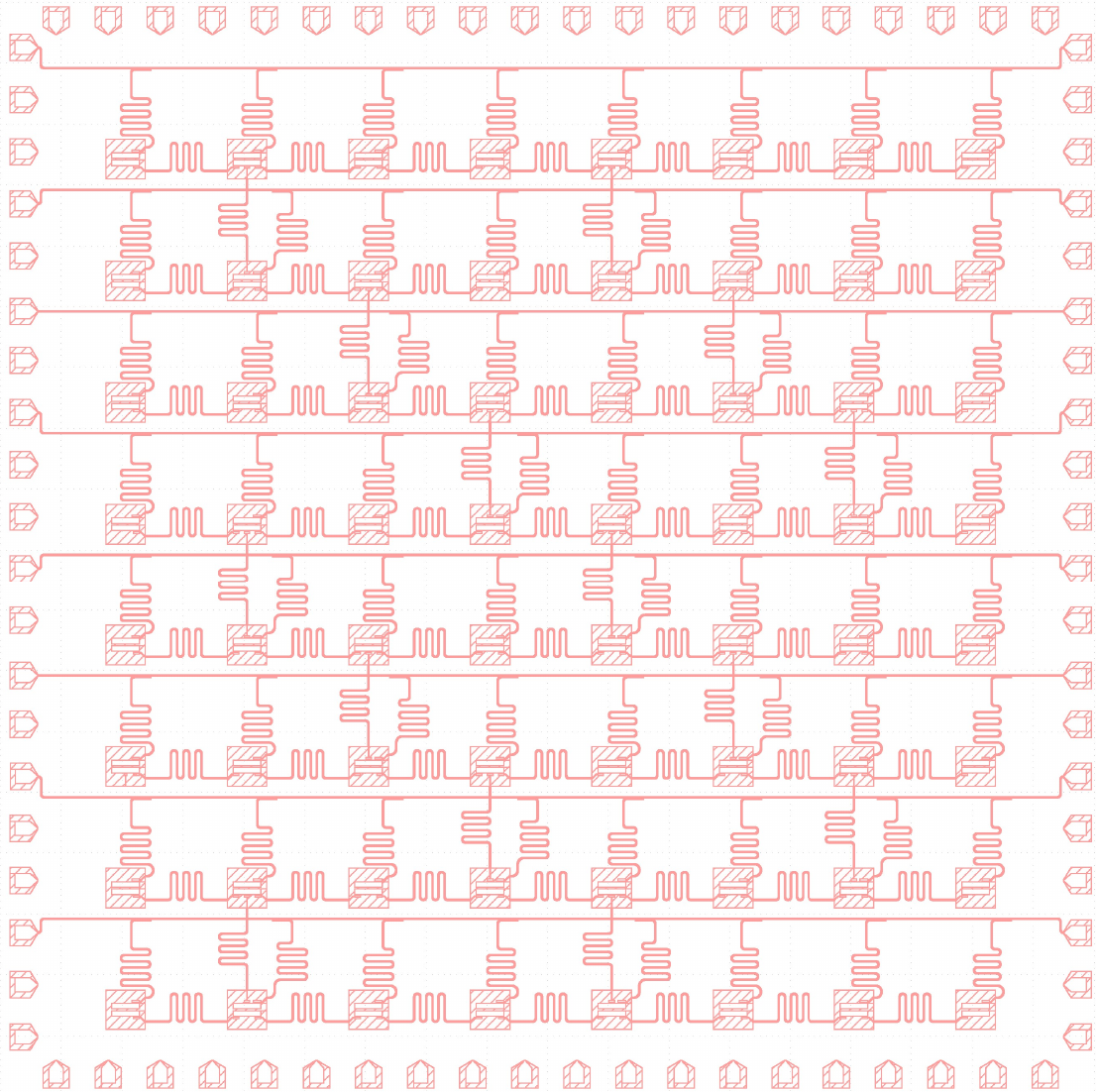}
        \caption{Post-optimization}
    \end{subfigure}
    \caption{The chip design process in practice, including topology design, equivalent circuit design, qubit layout design, automatic generation of readout cavity, automatic routing, and post-optimization. The post-optimization includes device mapping, process mapping, and chip layout adjustment.}
    \label{The chip design process in practice}
\end{figure}

\begin{figure}
    \centering
    \begin{subfigure}[b]{0.45\linewidth}
        \includegraphics[width=\linewidth]{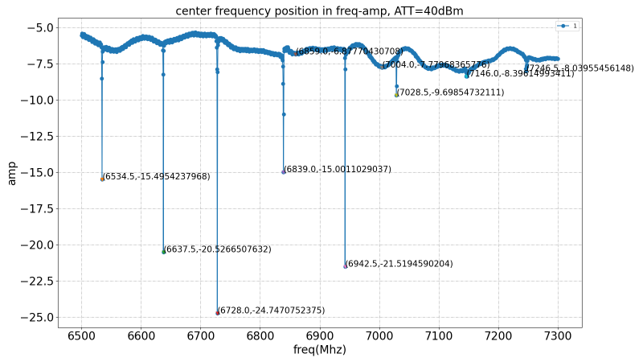}
        \caption{Bus E1F1}
    \end{subfigure}
    \hfill
    \begin{subfigure}[b]{0.45\linewidth}
        \includegraphics[width=\linewidth]{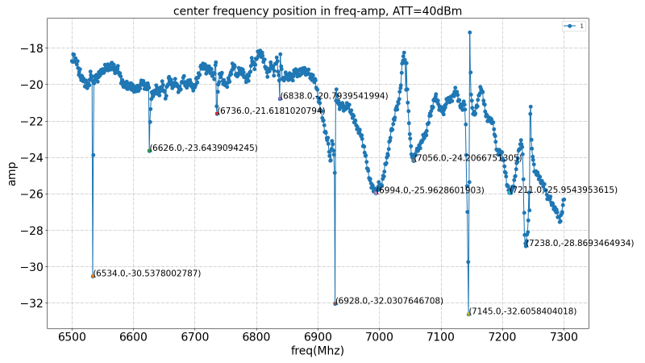}
        \caption{Bus E2F2}
    \end{subfigure}
    \begin{subfigure}[b]{0.45\linewidth}
        \includegraphics[width=\linewidth]{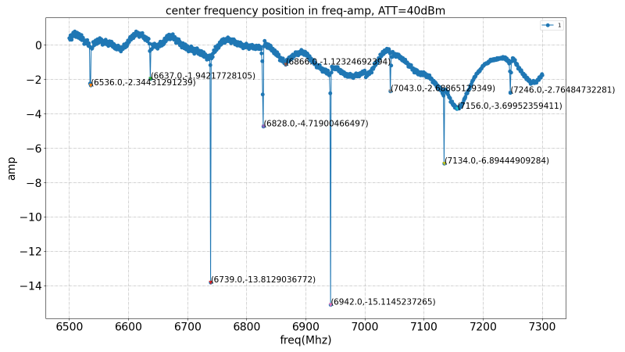}
        \caption{Bus E4F3}
    \end{subfigure}
    \hfill
    \begin{subfigure}[b]{0.45\linewidth}
        \includegraphics[width=\linewidth]{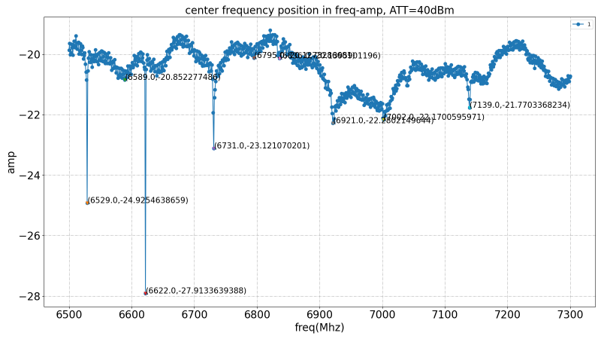}
        \caption{Bus E5F4}
    \end{subfigure}
    \begin{subfigure}[b]{0.45\linewidth}
        \includegraphics[width=\linewidth]{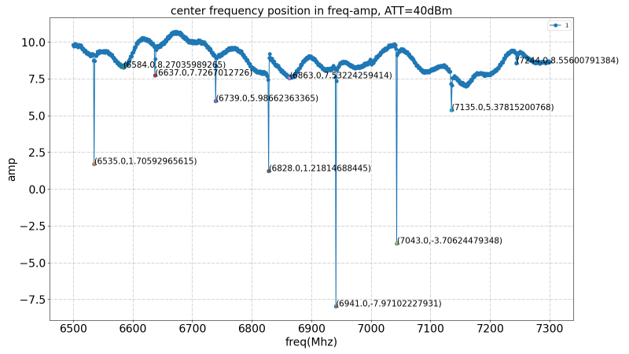}
        \caption{Bus B16C1}
    \end{subfigure}
    \hfill
    \begin{subfigure}[b]{0.45\linewidth}
        \includegraphics[width=\linewidth]{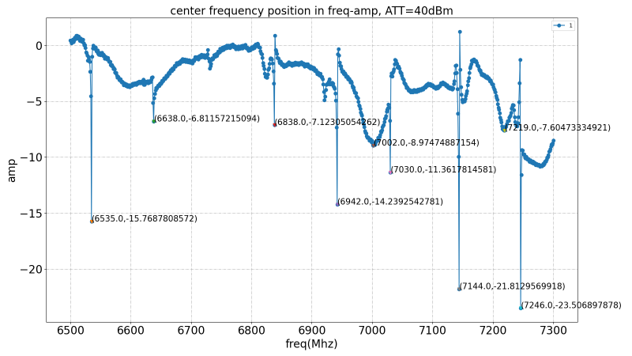}
        \caption{Bus B17C2}
    \end{subfigure}
    \begin{subfigure}[b]{0.45\linewidth}
        \includegraphics[width=\linewidth]{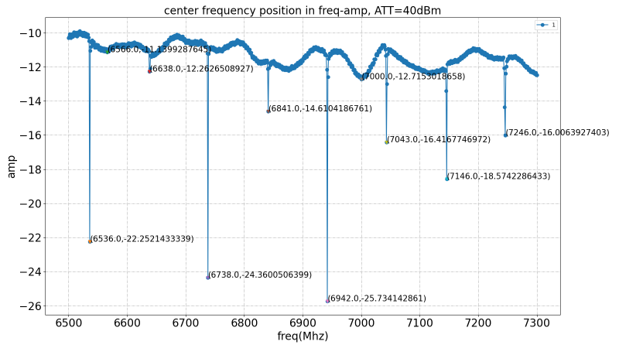}
        \caption{Bus B18C4}
    \end{subfigure}
    \hfill
    \begin{subfigure}[b]{0.45\linewidth}
        \includegraphics[width=\linewidth]{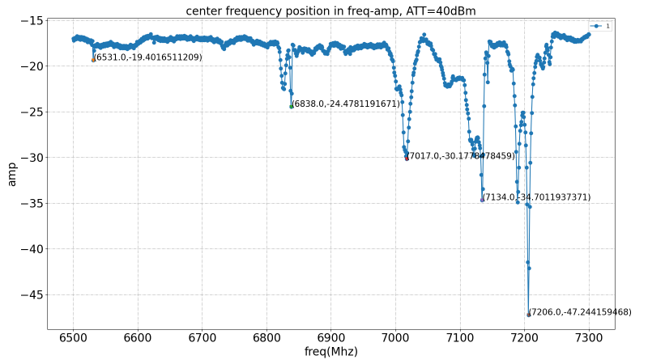}
        \caption{Bus B19C5}
    \end{subfigure}
    \caption{The chip test results.}

\end{figure}

The chip designed by EDA-Q was evaluated, and the Table \ref{tab:chip_parameters} displays the test data for certain qubits. The B18-C4 bus has an average T1 measurement value of 26.98 microseconds, while the E1-F1 bus has an average T1 measurement value of 66.18 microseconds. The E4-F3 bus recorded a qubit T1 value of 64.18 microseconds, while the B16-C1 bus recorded a qubit T1 value of 39.95 microseconds. Frequency sweeping is conducted on every bus, and the frequency of the readout cavity is measured, as illustrated in the Figure 13. The figure illustrates that the survival rate of the readout cavity on bus E4F3 and B18C4 reached 100\%.  Additionally, at least 6 read cavities with good quality were observed in bus E1F1, E2F2, B16C1, and B17C2. Some qubits and readout cavities test data have not been effectively tested due to the extensive impact of chip manufacturing process and measurement and control. However, this aspect is not the main focus of this paper.  Based on the current test results, EDA-Q demonstrates the capability to design efficient quantum chips.

\begin{table*}
	\centering
	\small
	\renewcommand{\arraystretch}{0.5}
	\caption{Chip Parameter}
	\label{tab:chip_parameters}
	\begin{tabular*}{\textwidth}{@{\extracolsep{\fill}}cccccccccc}
		\toprule
		Bus    & Qubit & Q1   & Q2    & Q3   & Q4    & Q5   & Q6   & Q7    & Q8    \\ \midrule
		B18-C4 & wr  & 6539 & 6642  & 6741.5 & 6844  & 6946 & 7046 & 7148.5 & 7249  \\
		& wq  & 4335 & 4200  & 4400   & 4160  & 4340 & 4180 & 4180   & 4125  \\
		& T1  & 25.71 & 28.84 & 34.05  & 33.19 & 18.55 & 14.34 & 21.7  & 39.44 \\ \midrule
		E2-F2   & wr  & 6535 & 6627.8 & 6736  & 6839  & 6929 & 7043 & 7147  & 7239  \\
		& wq  & \textbackslash &  \textbackslash  & \textbackslash & \textbackslash & \textbackslash & \textbackslash & \textbackslash & \textbackslash \\
		& T1  & \textbackslash & \textbackslash & \textbackslash & \textbackslash & \textbackslash & \textbackslash &  \textbackslash & \textbackslash \\ \midrule
		E1-F1   & wr  & 6534.5 & 6637.5 & 6728  & 6839  & 6942.5 & 7031 & 7148  & 7249  \\
		& wq  & 4520 & 4210  & 4350   & 4290  & 4780 & 4160 & 4310   & 4190  \\
		& T1  & 41.8 & 60    & 76.25  & 54.5  & 68.93 & 90   & 67.93  & 70    \\ \midrule
		E4-F3   & wr  & 6538.9 & 6641 & 6741  & 6830.2 & 6945.1 & 7046.9 & 7136 & 7249.2 \\
		& wq  & 4190 & 4480 & 4270  & 4390   & 4100   & 4380   & 4160 & 4380   \\
		& T1  & 22.9 & 20.3 & 123.2 & 50.8   & 105.7  & 78.8   & 101.2 & 10.5  \\ \midrule
		B19-C5  & wr  & 6531 & 6626.4 & 6735  & 6841.9 & \textbackslash & \textbackslash & \textbackslash & \textbackslash  \\
		& wq  & \textbackslash & \textbackslash & \textbackslash & \textbackslash  & \textbackslash & \textbackslash & \textbackslash & \textbackslash \\
		& T1  & \textbackslash & \textbackslash & \textbackslash & \textbackslash & \textbackslash & \textbackslash & \textbackslash & \textbackslash \\ \midrule
		B17-C2  & wr  & 6534.8 & 6637.7 & 6731.6 & 6837.6 & 6942  & 7021.5 & 7145.5 & 7249.3 \\
		& wq  & \textbackslash & \textbackslash  & \textbackslash & \textbackslash & \textbackslash  & \textbackslash & \textbackslash & \textbackslash \\
		& T1  & \textbackslash & \textbackslash  & \textbackslash & \textbackslash & \textbackslash & \textbackslash & \textbackslash & \textbackslash \\ \midrule
		B16-C1  & wr  & 6535 & 6637   & 6739   & 6828   & 6941  & 7043   & 7135   & 7246   \\
		& wq  & 4500 & 4170   & 4420   & 4210   & 4233  & 4140   & 4300   & 4130   \\
		& T1  & 31.96 & 40    & 42.6   & 40     & 39.77 & 45.24  & 50     & 30     \\ \midrule
		E5-F4   & wr  & 6532 & 6625   & 6734   & 6838   & 6924  & 7042   & 7139   & \textbackslash \\
		& wq  & 4290 & 4410   & 4100   & 4410   & 4160  & 4350   & \textbackslash & \textbackslash \\
		& T1  & 46.44 & 9.36  & 22.28  & 40     & 13.95 & 40     & \textbackslash  & \textbackslash \\
		\bottomrule
	\end{tabular*}
\end{table*}

\section{Conclusion and Expectation}
EDA-Q is a pioneering automated design tool for superconducting quantum chips that supports the entire design process, including topology to simulation.   It incorporates device mapping and fabrication process mapping capabilities into quantum chip design tools for the first time.   EDA-Q utilizes a software architecture that is adaptable, extensible, and compatible with the constantly evolving technological environment.   EDA-Q provides straightforward and flexible command interfaces that ensure a user-friendly experience, even in intricate application scenarios.   For our future works, we are actively collaborating with different chip fabrication manufacturers in order to establish a standardized set of industry guidelines for fabrication process libraries.   In addition, we are enhancing the system's algorithm library, specifically in the fields of topological design, automated routing, and device mapping.   Researchs on algorithms for these fields has the potential to greatly improve EDA-Q's capabilities.   Continual efforts are being made to expand the device and simulation libraries in order to cater to the varied requirements of users.

The source code for this project is available under the [GPL-3.0 license] at [https://github.com/Q-transmon-xmon/EDA-Q].

\end{document}